\documentstyle[prd,preprint,aps,epsfig,floats]{revtex}
\begin{document}

\tightenlines
\input epsf.tex
\def\DESepsf(#1 width #2){\epsfxsize=#2 \epsfbox{#1}}
\draft
\thispagestyle{empty}
\preprint{\vbox{\hbox{OSU-HEP-01-08} \hbox{UMD-PP-02-015}
\hbox{October 2001}}}

\title{\Large \bf $L_e+L_{\mu}-L_{\tau}-L_s$ Symmetry
and a Mixed 2+2 Scenario for Neutrino Oscillations}

\author{\large\bf K.S. Babu$^{(a)}$ and R.N. Mohapatra$^{(b)}$}

\address{(a) Department of Physics, Oklahoma State University\\
Stillwater, OK 74078, USA \\}

\address{(b) Department of Physics, University of Maryland\\
College Park, MD 20742, USA}

\maketitle

\thispagestyle{empty}

\begin{abstract}

Recent results from  SuperKamiokande and SNO experiments have set severe
constraints on possible mixings of a light sterile neutrino, $\nu_s$, with the
three active species required for a simultaneous explanation of the solar,
atmospheric and LSND neutrino oscillation data. A consistent scheme has
emerged from a global analysis of the data wherein two of the neutrinos
are nearly degenerate with a mass of order 1 eV, that mix significantly
with the two lighter states.  We present
realizations of such a mixed 2+2 oscillation scenario based on
$L_e+L_\mu-L_\tau-L_s$ symmetry ($L_i$ stands for the $i$th lepton number).
Breaking of of this lepton number symmetry by a small mass term for $\nu_s$
leads to the required large mixings for both the atmospheric and the solar
neutrino oscillations.  Sum rules for the neutrino oscillation parameters
are derived within this scheme, and are shown to be consistent with present data.
These models predict $U_{e3} \simeq 0.02-0.03$, which can serve as a test
of this idea.  We also present gauge models based on mirror extensions of the
Standard Model that naturally lead to a light sterile neutrino with the required
mixing pattern.

\end{abstract}

\newpage
\section{Introduction}

Neutrino oscillation data from three different classes
of experiments, viz., atmospheric \cite{atm},
solar \cite{cl,solar,sage,gallex,sno},
and LSND \cite{lsnd}, provide conclusive evidence
in favor of non-zero neutrino masses and mixings and thus for
physics beyond the Standard Model. In order to understand the specific
nature of new physics, one has to determine the detailed
pattern of mixings and masses that fit various observations. As more
and more experimental data become available, these details are getting
less fuzzy \cite{review} and perhaps more interestingly, some simple
possibilities are being either heavily disfavored or eliminated.
In this note, we study the nature of new physics and possible new symmetries
implied by the three pieces of observation taken simultaneously.

Broadly speaking, since the three different types of experiments
are sensitive to three different scales of oscillation lengths, or
equivalently three different values of mass splittings $\Delta m^2$,
a simultaneous explanation of all
data requires that we go beyond the conventional framework of
three known light neutrinos $(\nu_e,\nu_{\mu},\nu_{\tau})$ and
postulate the existence of a fourth light sterile neutrino, $\nu_s$.
This is the four neutrino picture that we will seek to
understand in this paper. Needless to say that the need for the sterile
neutrino will be somewhat tentative until the LSND results are confirmed by
new experiments such as the MiniBOONE experiment at Fermilab.

There are three classes of models that can potentially explain
all the data:  (i) 2+2 scenario,
(ii) 3+1 scenario and (iii) the mixed 2+2 scenario.

\vspace*{0.1in}
\noindent{\bf 2+2 scenario:}
\vspace*{0.1in}

In this scenario \cite{2+2,pakvasa}, two of the neutrino are nearly degenerate
and have masses approximately equal to
$\sqrt{\Delta m^2_{LSND}} \sim 1$ eV, while the remaining two states
are much lighter. The two heavy eigenstates
are generally chosen to consist mainly of the near-maximally mixed weak
eigenstates $\nu_{\mu}$ and $\nu_{\tau}$ with
their mass difference squared of order $\Delta m^2_{atm}\simeq 3\times 10^{-3}$
eV$^2$. Clearly this is designed to explain the atmospheric neutrino
data via $\nu_{\mu}-\nu_{\tau}$ oscillation. The light eigenstates are
assumed to contain mostly
$\nu_e$ and $\nu_s$ with a mass difference squared equal to the solar
splitting $\Delta
m^2_{\odot}$. There is a small mixing between the two
pairs which provides an explanation of the LSND results.

\vspace*{0.1in}
\noindent{\bf 3+1 scenario:}
\vspace*{0.1in}

In this scheme \cite{3+1}, the three active neutrinos
$\nu_{e,\mu,\tau}$ have masses $\leq \sqrt{\Delta m^2_{atm}}\simeq 0.05$ eV,
while $\nu_s$ is
heavier with a mass of order $\sqrt{\Delta m^2_{LSND}} \sim 1$ eV. The solar and
atmospheric neutrino oscillations then involve only the active neutrinos
and LSND data is explained by indirect oscillations involving the sterile
neutrino\cite{babu}.  The goodness of fit to the LSND data, while acceptable,
is not great \cite{grimus}.
Only certain values of the $\Delta
m^2_{LSND}$ are allowed and that too at the 95\% confidence level in this
scheme.

\vspace*{0.1in}
\noindent{\bf Mixed 2+2 scenario:}
\vspace*{0.1in}

This scenario is a variation of the 2+2 scheme which is necessitated by
recent solar neutrino data from SNO experiment in conjunction with
SuperKamiokande data.
Prior to the SNO results, the 2+2 scenario where solar neutrino
oscillations were assumed to involve only the $\nu_e$ and $\nu_s$ was
considered to give a good fit to the data although the fit to the
energy distribution was not as good as the fit to the rates.
(The 3+1 scenario gave a better fit to the energy spectrum as well as the
rates but could fit LSND data only at the 95\% CL.)

The SNO results \cite{sno} have had a major impact on these
conclusions. It
showed that there is a clear gap between the neutrino flux
measured by the SuperKamiokande experiment and the charged current
flux measured by the SNO experiment. If the solar neutrinos oscillate only
to the sterile neutrinos as is assumed in the original 2+2 scenario, the
fluxes measured by SNO and SuperKamiokande should be equal. The observed
gap in the flux would appear to exclude the 2+2 scenario.  The mixed 2+2
scenario circumvents this problem.  In this
scenario, solar $\nu_e$ is assumed to oscillate into a linear combination
of $\nu_s$ and $\nu_{\mu,\tau}$.  This would allow an understanding of the
gap in the flux measured at SuperKamiokande and SNO, since the $\nu_{\mu,\tau}$
components in solar neutrinos will contribute to neutral current cross
section at SuperKamiokande, but will not play any role in the charged
current cross section at SNO.
The challenge then is to achieve such a mixed oscillation scenario
without upsetting the other constraints from accelerator data on the
one hand and the atmospheric neutrino oscillation data on the other.
Atmospheric neutrino data do set severe constraints on such a mixed
scheme since that prefers $\nu_\mu-\nu_\tau$ oscillations with only
a limited amount of $\nu_s$ component allowed.

In a recent paper, Gonzales-Garcia et. al. \cite{concha}
have studied this issue in quantitative detail.  They proposed a $4\times 4$
neutrino mixing matrix where $U_{e3} \simeq U_{e4} \simeq 0$ was assumed in
order to satisfy bounds from reactor neutrino data, especially from the CHOOZ
\cite{chooz} and PALO-VERDE \cite{paloverde} experiments.  Here the two heavy
eigenstates with nearly degenerate mass of order 1 eV are denoted by
$\nu_3$ and $\nu_4$, while  $\nu_1$ and $\nu_2$ denote the lighter eigenstates.
There is substantial mixing between the heavy and the light sectors.  Solar
neutrinos can oscillate into $\nu_{\mu,\tau}$ with a probability
proportional to $(1-|U_{s1}|^2-|U_{s2}|^2)$ where $U_{s1}$ and $U_{s2}$
parametrize the non-negligible sterile neutrino component in the two light
eigenstates $\nu_1$ and $\nu_2$.  This helps reconcile the SNO and the
SuperKamiokande solar neutrino data.  On the other hand, this mixing
pattern allows atmospheric $\nu_{\mu}$'s to oscillate into sterile
neutrinos with a probability of $2(|U_{s1}|^2+|U_{s2}|^2)(1-|U_{\mu 3}|^3-
|U_{\mu 4}|^2)$. The limit on the
sterile neutrino fraction in the atmospheric data therefore tends to
reduce the amount of the needed active neutrino fraction in the solar neutrino
data. A delicate balance is needed and the authors of Ref.\cite{concha}
found a fit to both the solar (including SNO data) and atmospheric data
for a range of values of the mixing angles consistent with all other
observations. This mixing pattern
is an interesting way to accommodate the LSND data for a wider range of
$\Delta m^2_{LSND}$ than the 3+1 scenario and in fact this delicate
balance means that this mixing pattern can be tested once the data
improves and as new experiments give results.

In this paper, we explore theoretical scenarios that can lead to mixed 2+2
scenarios with mixing patterns close to the one just described. We find
that this can happen if there is an approximate leptonic symmetry
$L_e+L_{\mu}-L_{\tau}-L_s$ in the Majorana mass matrix for the four
neutrinos in the context of a seesaw  mechanism for neutrino masses.
If this symmetry is broken by a mass term for $\nu_s$, the large
mixing angles needed for both solar and atmospheric neutrino oscillations
will result.  We show that this scenario predicts certain sum rules
relating neutrino oscillation parameters which are in agreement with
the data currently.
We then show that if the Standard Model is duplicated to have a mirror
sector, the lightest mirror neutrino can play the role of the
light sterile neutrino with the required mixing properties.
The desired leptonic symmetry emerges at low
energies, if the original model at high energies has the symmetry
$(L_e+L_{\mu}-L_{\tau})\times (L'_e+L'_{\mu}-L'_{\tau})$ (here prime denotes
the symmetries of the mirror sector).

We have organized this paper as follows: In section II, we describe the
phenomenology of the mass matrix with the approximate leptonic symmetry
described above; in section III, we study the phenomenological
implications and numerical fits to data for this mass matrix
and derive certain sum rules for oscillation parameters. In section
IV, we present a gauge model based on the existence of
a mirror sector that leads naturally to a light sterile neutrino
with the desired mixing pattern. In section V we conclude.

\section{Approximate $L_e+L_\mu-L_\tau-L_s$ Symmetry and
the Mixed (2+2) Neutrino Oscillation Scheme}

Consider the four--neutrino system ($\nu_e,\nu_\mu,\nu_\tau,\nu_s$).
Let the mass matrix
be of the form:

\begin{eqnarray}
M_\nu = \left(\matrix{ 0 & 0 & \epsilon_2 & \epsilon_1 \cr
0 & 0 & 1 & a \cr
\epsilon_2 & 1 & 0 & 0 \cr
\epsilon_1 & a & 0 & \delta}\right)m_0~.
\end{eqnarray}

\noindent This matrix has an approximate $U(1)$ symmetry which can be identified
as $L_e+L_\mu-L_\tau-L_s$ where $L_i$ stands for the $i$th lepton number.  This
$U(1)$ symmetry is not exact, it is broken by the $\nu_s$ mass term $(\delta m_0)
(\nu_s\nu_s)$ in Eq. (1).  This breaking will turn out to be small,
in fact $\delta$ will be the smallest entry in Eq. (1).  In the limit
$\delta \rightarrow
0$, $M_\nu$ of Eq. (1) will collapse effectively to a $2 \times 2$ matrix, with
the four neutrino states forming two Dirac fermions.  The mass matrix in the
$\delta \rightarrow 0$ is given by

\begin{eqnarray}
M_\nu^0 = \left(\matrix{\nu_e & \nu_\mu}\right)^T~~
\left(\matrix{\epsilon_1 & \epsilon_2 \cr a & 1}\right)m_0~~\left(\matrix{
\nu_s \cr \nu_\tau}\right)~.
\end{eqnarray}

\noindent Note that due to the $L_e+L_\mu-L_\tau-L_s$ symmetry,
mixing occurs only between $(\nu_e,~\nu_\mu)$ and $(\nu_\tau,~\nu_s)$ in
this limit.
We shall be interested in the case where $a \sim 1 \gg \epsilon_1,
\epsilon_2$.
The two nonzero mixing angles are then
\begin{eqnarray}
\tan\theta_{s\tau} &\simeq& a ~,\nonumber \\
\theta_{e\mu} &\simeq& {\epsilon_1 a + \epsilon_2 \over 1+a^2} ~.
\end{eqnarray}
The two Dirac neutrinos will have masses given by
\begin{eqnarray}
m_h &\simeq& \sqrt{1+a^2}~ m_0 ~,\nonumber \\
m_l &\simeq& {|\epsilon_1-\epsilon_2 a| \over \sqrt{1+a^2}}~m_0~.
\end{eqnarray}

\noindent The mass splitting relevant for the LSND experiment is
$\Delta m^2_{LSND} \simeq m_h^2-m_l^2 \simeq (1+a^2)m_0^2$.
We shall choose $m_h \simeq 1$ eV so as to explain the LSND experiment,
along with $\epsilon_{1,2} \simeq (2-3) \times 10^{-2}$ to fulfill
the LSND mixing angle requirement. At this stage, no mass splitting
other than that for LSND is induced.  We can allow for the
possibility that either $\epsilon_1$
or $\epsilon_2$ is zero due to some flavor symmetry.  These special
cases will reduce the number of parameters by one.
We shall keep both terms to be nonzero to be general, but in Sec. III,
we shall also discuss these two special cases.

Now, let us include the effects of $L_e+L_\mu-L_\tau-L_s$ symmetry breaking
through the mass term $\delta$ (the (4,4) entry of Eq. (1)).
This mass term, which breaks the symmetry rather economically,
will serve several purposes.
It will induce two more mass splittings, to be identified with the
atmospheric mass splitting ($\Delta m^2_{atm}$) and the
solar mass splitting ($\Delta m^2_{\odot}$).  At the same time the $\delta$ term
will lead to maximal or near maximal mixing between the two would be
Dirac states
both in the atmospheric neutrino sector and in the solar neutrino sector.

On the theoretical side, we envision that the global $L_e+L_\mu-L_\tau-L_s$
symmetry is broken by some high scale physics.  For example, quantum
gravity is suspected
to break all global symmetries, so that could be the source of the
$\delta$ term.
It is quite natural to assume that the symmetry breaking effects show up first
in the $\nu_s$ mass term, which is a complete singlet of the Standard Model.
In fact, in the explicit gauge models that we have constructed (see
Sec. IV),
$\delta$ arises through quantum gravity, and the analogous symmetry breaking
effects are negligible in all other entries of Eq. (1).

Including the $\delta$ term, the eigenvalues of $M_\nu$ of Eq. (1) can be
computed in the approximation $1 \sim a \gg \epsilon_{1,2} \sim \delta$.
Neglecting quadratic terms in $\epsilon_{1,2}$ and $\delta$, these masses
are:
\begin{eqnarray}
m_4 &\simeq& \left(\sqrt{1+a^2} + {a^2 \delta \over 2(1+a^2)}\right)m_0~,
\nonumber \\
m_3 &\simeq& \left(-\sqrt{1+a^2} + {a^2 \delta \over 2(1+a^2)}\right)m_0~,
\nonumber \\
m_2 &\simeq& \left({\delta + X \over 2(1+a^2)}\right) m_0 ~,\nonumber \\
m_1 &\simeq& \left({\delta - X \over 2(1+a^2)}\right)m_0~.
\end{eqnarray}

\noindent Here we have defined
\begin{equation}
X \equiv \sqrt{\delta^2+4(1+a^2)(\epsilon_1-a \epsilon_2)^2}~
\end{equation}
for convenience.  We have tacitly assumed all parameters of $M_\nu$ to
be real for simplicity.  For $\delta$ positive,
we have $m_1 \leq m_2 \leq m_3 \leq m_4$.  If $\delta$ is negative, we
can rearrange the labels, $m_1 \leftrightarrow m_2$ and $m_3
\leftrightarrow m_4$,
so that the hierarchy $m_1 \leq m_2 \leq m_3 \leq m_4$ is maintained.

From Eq. (5) the three relevant mass splittings are found to be
\begin{eqnarray}
\Delta m^2_{LSND} &\simeq& m_4^2-m_2^2 \simeq (1+a^2)~m_0^2 ~,\nonumber \\
\Delta m^2_{atm} &\simeq& m_4^2-m_3^2 \simeq \left({2 a^2 \delta \over
1+a^2}\right)m_0^2~,
\nonumber \\
\Delta m^2_{\odot} &\simeq& m_2^2-m_1^2 \simeq (2\delta X)~ m_0^2~.
\end{eqnarray}

The leptonic mixing matrix $U$ is given by (to linear order in $\delta$
and $\epsilon_{1,2}$)

{\tiny
\begin{eqnarray}
U=
\left(\matrix{
-{\sqrt{2}\sqrt{1+a^2}(\epsilon_1-a\epsilon_2) \over
\sqrt{X(X-\delta)}} &
{\sqrt{2}\sqrt{1+a^2}(\epsilon_1-a\epsilon_2) \over
\sqrt{X(X+\delta)}} &
-{(a\epsilon_1+\epsilon_2) \over \sqrt{2}(1+a^2)} &
{(a\epsilon_1+\epsilon_2) \over \sqrt{2}(1+a^2)} \cr
-{\sqrt{2} [(a\epsilon_1+\epsilon_2)(\epsilon_1-
a\epsilon_2) + {a \delta \over 2(1+a^2)}(\delta-X)] \over
\sqrt{1+a^2}\sqrt{(X(X-\delta)}} &
-{\sqrt{2}  [(a\epsilon_1+\epsilon_2)(\epsilon_1-
a\epsilon_2) + {a \delta \over 2(1+a^2)}(\delta+X)] \over
\sqrt{1+a^2}\sqrt{(X(X+\delta)}} &
-{1 \over \sqrt{2}} - {a^2 \delta \over 4\sqrt{2}(1+a^2)^{3/2}} &
{1 \over \sqrt{2}} - {a^2\delta \over 4\sqrt{2}(1+a^2)^{3/2}} \cr
-{a\over \sqrt{2}\sqrt{1+a^2}}\sqrt{{X-\delta \over X}} &
-{a\over \sqrt{2}\sqrt{1+a^2}}\sqrt{{X+\delta \over X}} &
{1 \over \sqrt{2} \sqrt{1+a^2}} + {3 a^2 \delta \over 4 \sqrt{2}(1+a^2)^2} &
{1 \over \sqrt{2}\sqrt{1+a^2}} - {3 \delta a^2 \over 4\sqrt{2}(1+a^2)^2} \cr
{1 \over \sqrt{2}\sqrt{1+a^2}}\sqrt{{X-\delta \over X}} &
{1 \over \sqrt{2}\sqrt{1+a^2}}\sqrt{{X+\delta \over X}} &
{a \over \sqrt{2}\sqrt{1+a^2}} - {a \delta (4+a^2) \over 4\sqrt{2}(1+a^2)^2} &
{a \over \sqrt{2}\sqrt{1+a^2}} + {a \delta (4+a^2) \over 4\sqrt{2}(1+a^2)^2}
}\right)
\end{eqnarray}
}

\noindent Here the entries in the first row are $U_{ei}$, $i=1-4$, the ones
in the second row are $U_{\mu i}$, the third row entries are $U_{\tau i}$ and
the last row entries are $U_{si}$.  Here we have used
the definition $\nu_\alpha = \sum_{k=1}^4
U_{\alpha k}\nu_k$, where $a=(e,\mu,\tau,s)$ denote the four flavors
and $k=1-4$ denote the mass eigenstates.

\section{Numerical Fits}

With the mass splittings and the mixing matrix entries in hand, we can now
confront the model with oscillation data from LSND, solar neutrino
and atmospheric neutrino experiments.

\subsection{LSND experiment}

From the expression for $\Delta m^2_{LSND}$ given in Eq. (7), we see that
$(1+a^2)m_0^2 \simeq (0.2-6)$ eV$^2$ in order to explain the positive
results seen by the LSND collaboration.  The ($\nu_\mu-\nu_e$) oscillation
probability relevant for LSND is given by
\begin{equation}
P_{\nu_\mu\rightarrow \nu_e}(LSND) \simeq 4\left|U_{e3}^*U_{\mu 3}+
U_{e4}^*U_{\mu 4}\right|^2\sin^2\left({\Delta m^2_{LSND} L\over 4E}\right)~,
\end{equation}
where we have ignored the contribution proportional to the smaller
$\Delta m^2_\odot$
and $\Delta m^2_{atm}$
and assumed that $m_3 \simeq m_4$.  Comparing Eq. (9) with the matrix elements
of Eq. (8), we make the identification
\begin{equation}
\sin^22\theta_{LSND} \approx  4\left|U_{e3}^*U_{\mu 3}+
U_{e4}^*U_{\mu 4}\right|^2 \simeq 4 \left[{a \epsilon_1 + \epsilon_2 \over (1+a^2)}
\right]^2~.
\end{equation}
Here $\sin^22\theta_{LSND} \simeq 3 \times 10^{-3}$ is the mixing parameter
usually quoted in the two flavor oscillation analysis.  The
experimental observation can be explained
by choosing $(a \epsilon_1+\epsilon_2)/(1+a^2) \simeq 0.03$.  Note that for this
analysis, the breaking of $L_e+L_\mu-L_\tau-L_s$ symmetry is not significant,
we could have obtained identical results by performing the two flavor oscillation
using Eq. (2) in the exact limit of this symmetry.

Note that the effective mixing parameter $U_{e3}$, normally discussed in the
three neutrino oscillation scenario, is given in our model
by $|U_{e3}^{eff}|^2 = (|U_{e3}|^2+|U_{e4}|^2)$ since $m_3 \simeq m_4$.
Thus $|U_{e3}^{eff}| \simeq \theta_{LSND} \simeq (0.02-0.03)$.  This
is a definite prediction of the model that can be used as one of its
tests.

\subsection{Solar and atmospheric neutrino oscillations}

We shall follow the global analysis of solar and atmospheric neutrino
oscillation
data carried out in Ref. \cite{concha} including the recent SNO
results.  In that
paper it was assumed that $U_{e3} = U_{e4} = 0$.  This  approximation holds
to a high degree of accuracy in our case,
since $U_{e3} \simeq U_{e4} \simeq \theta_{LSND}/\sqrt{2} \simeq 0.02$.
Such a small mixing of $\nu_e$ with the heavier mass eigenstates is
insignificant
for the analysis of solar and atmospheric neutrino oscillations.

Consider first the atmospheric neutrino oscillations.  The parameter space
is determined by $\{\Delta m^2_{atm},~|U_{\mu 3}|^2/(|U_{\mu
3}|^2+|U_{\mu 4}|^2),~
(|U_{s1}|^2+|U_{s2}|^2),~
(|U_{\mu 1}|^2+U_{\mu 2}|^2)\}$.  Here the mass splitting is given in Eq. (7),
which we shall choose to be $\Delta m^2_{atm} \simeq (1-6) \times
10^{-3}$ eV$^2$.
The mixing parameter $|U_{\mu 3}|^2/(|U_{\mu 3}|^2+|U_{\mu 4}|^2) \simeq
1/2$
from Eq. (8). This is the leading parameter that
controls the disappearance of $\nu_\mu$ through oscillations into either
$\nu_\tau$
or $\nu_s$.  Our model prediction
agrees quite well with the results of the global analysis \cite{concha}.
In particular, our model predicts the deviation of this parameter from $1/2$ to
be extremely small, of order 1\%.
The parameter
$(|U_{s1}|^2+|U_{s2}|^2) \simeq 1/(1+a^2)$ parametrizes the projection of
the sterile
neutrino component in the atmospheric neutrino oscillations.  When this
parameter is
equal to 1, we have pure $\nu_\mu-\nu_\tau$ oscillations, while if it were zero
we have pure $\nu_\mu-\nu_s$ oscillations.  The atmospheric neutrino data
prefers
$\nu_\mu$ oscillations into mostly $\nu_\tau$, but significant sterile
component is also
allowed.  When combined with the solar neutrino analysis, which prefers this
parameter to be small rather than large, Ref. \cite{concha} obtains
$(|U_{s1}|^2+|U_{s2}|^2) \simeq 0.21-0.5$ as preferred \cite{concha}.
That fixes the parameter $a = (1-2)$.  Lastly,  the mixing parameter
$(|U_{\mu 1}|^2 + |U_{\mu 2}|^2) \simeq  \theta_{LSND}^2 +
a^2\delta^2/(1+a^2)^3$,
as can be seen with a little algebra.
Numerically this is about 0.001, which is too small to be of significance in
atmospheric oscillations.  Thus, our model corresponds to the restricted case
of $(|U_{\mu 1}|^2 + |U_{\mu 2}|^2) =0$ studied in
Ref. \cite{concha,fogli}.
This scenario gives a reasonable global fit with a goodness of fit
(GOF) quoted to
be 59\% \cite{concha}.  Actually, within our scheme this GOF will be somewhat
better since one of the parameter $|U_{\mu 3}|^2/(|U_{\mu 3}|^2+|U_{\mu
4}|^2)$
is fixed to be $1/2$, which is very close to the central value of the
experiments.

Thus, a good fit to atmospheric neutrino oscillation within our
scheme requires $a \simeq (1-2)$,
and $\delta = (2\times 10^{-3}-3 \times 10^{-2})$, if we make use of
$\Delta m^2_{LSND}
\simeq (0.5-1)$ eV$^2$.  The parameter $(a\epsilon_1+\epsilon_2)$ lies in
the range
$(0.06-0.15)$, obtained from fitting the LSND mixing angle.  We see that
$\delta, \epsilon_{1,2} \ll 1$ while $a\sim 1$, consistent with our
approximations.
Note also that while $\delta$ is somewhat smaller than $\epsilon_{1,2}$, it is
not much smaller.  In particular, $\delta$ may be comparable to $(\epsilon_1-
a\epsilon_2)$ for a wide range of parameters.  This observation will be relevant
for the solar neutrino mixing angle prediction.

Turning to solar neutrinos,
the parameter space is characterized by ($\Delta
m^2_{\odot},~|U_{e2}|^2/(1-|U_{e2}|^2)$,
$(|U_{s1}|^2+|U_{s2}|^2)$).  The mass splitting is given in Eq. (7).  The first
mixing parameter is recognizable as $\tan^2\theta_\odot$ that is usually used
in the two flavor oscillation analysis for $\nu_e$ disappearance.  The second
mixing parameter specifies the amount of sterile neutrinos in solar neutrino
oscillations.  When $(|U_{s1}|^2+|U_{s2}|^2)=0$, $\nu_e$ oscillates only to
active species, when this parameter is 1, $\nu_e$ oscillates into a pure sterile
state.  Solar neutrino data prefers $\nu_e$ oscillating predominantly to an
active species, with some sterile admixture allowed.

From Eq. (8), we see that $\tan^2\theta_\odot \simeq (X-\delta)/(X+\delta)$.
If $\delta \ll X$, this angle will be maximal.  However, $\delta$ and $X$
may be comparable, so $\tan^2\theta_\odot$ can deviate significantly from
one.  Note that the deviation will be to lower values compared to one, a feature
that goes well with oscillation data.

\subsection{Sum rules for oscillation parameters}

Now we show that the model under study leads to an interesting sum rule
when the oscillation data from LSND, solar, and atmospheric neutrinos are
combined.  To see this, note that $(|U_{s1}|^2+|U_{s2}|^2) \simeq 1/(1+a^2)$
fixes the value of $a$.  The solar oscillation angle
$\tan^2\theta_\odot$ fixes the value of $\delta/X$, while the ratio
$\Delta m^2_\odot/\Delta m^2_{LSND}$ fixes the product $\delta X$.  The
parameters ($a,~\delta,~X$) are then completely fixed.  The sum rule arises
by examining the ratio $\Delta m^2_{atm}/\Delta m^2_{LSND} \simeq
2a^2\delta/(1+a^2)^2$, which only depends on the same set of parameters.
A simple calculation shows the sum rule to be
\begin{equation}
\Delta m^2_\odot \cos2\theta_\odot \simeq {
\left[\Delta m^2_{atm}\right]^2 \over 2 \Delta
m^2_{LSND}}{1 \over
(|U_{s1}|^2+|U_{s2}|^2)(1-|U_{s1}|^2-|U_{s2}|^2)^2}~.
\end{equation}

\noindent This sum rule is obeyed by current experiments rather well.  As an
example, let us choose $(|U_{s1}|^2+|U_{s2}|^2) = 0.25$, $\Delta
m^2_{atm} = 3 \times
10^{-3}$ eV$^2$, $\Delta m^2_{LSND} = 1$ eV$^2$, $\tan^2\theta_\odot = 0.5$.
Eq. (11) then predicts $\Delta m^2_\odot = 9.6 \times 10^{-5}$ eV$^2$.  This
value is nicely consistent with large angle solar neutrino oscillations.

Consider now the special case $\epsilon_1=0$, $\epsilon_2 \neq 0$,
which may be imposed by some
flavor symmetry.  In this case, there is an additional sum rule, which may
be taken to be a relation for the LSND mixing angle.  (This can be
seen by noting that $4(1+a^2)(\epsilon_1-a\epsilon_2)^2 \simeq X^2-\delta^2$.)
\begin{equation}
\theta_{LSND} \simeq {\Delta m^2_{atm} \over 4 \Delta m^2_{LSND}}{
(|U_{s1}|^2+|U_{s2}|^2)\tan 2\theta_\odot \over (1-|U_{s1}|^2-|U_{s2}|^2)^{3/2}}~.
\end{equation}

\noindent To see how this sum rule compares with experiment, let us take
$\Delta m^2_{atm} = 2 \times 10^{-3}$ eV$^2$, $\Delta m^2_{LSND} = 0.5$ eV$^2$,
$(|U_{s1}|^2+|U_{s2}|^2)= 0.7$, $\tan^2\theta_\odot = 0.75$.  From Eq. (11)
this choice will predict $\Delta m^2_\odot = 4.4 \times 10^{-4}$ eV$^2$, and
from Eq. (12), $\theta_{LSND} = 0.03$.  Both predictions are in good
agreement with observations.

In the opposite case, where $\epsilon_2=0$, $\epsilon_1 \neq 0$, there is
again a sum rule analogous to Eq. (12), given by
\begin{equation}
\theta_{LSND} \simeq {\Delta m^2_{atm} \over 4 \Delta m^2_{LSND}}{\tan 2\theta_\odot
\over (1-|U_{s1}|^2-|U_{s2}|^2)^{1/2}}~.
\end{equation}

\noindent In this case, $\theta_{LSND}$ tends to be small, $\theta_{LSND} \simeq
0.013$ for the same set of parameters as above.  So this special case appears
to be disfavored by present data, unless LSND mixing angle settles to a
much smaller value.

\section{Gauge Models}

The mass matrix of Eq. (1) can be obtained from extensions of the Standard Model
that incorporate the seesaw mechanism along with an approximate
$L_e+L_\mu-L_\tau-L_s$
symmetry.  In order to have a naturally light sterile state $\nu_s$ in the light
spectrum, we resort to the idea of mirror universe \cite{mirror,bere}.  In
this scenario, it is
assumed that there is a mirror sector to the Standard Model, which is identical
to our world in terms of its particle and force content. The symmetry
breakings can however be different\cite{bere}. The main advantage of this
extension
is that the existence of the extra gauge quantum numbers of the new
neutrino species makes it easier to understand the smallness of their masses.
For instance one can employ a type II seesaw mechanism, i.e., seesaw via
tiny VEV of a superheavy triplet Higgs \cite{type2}
instead of the heavy right handed Majorana
neutrinos in both sectors to make all active as well as sterile neutrinos light.
The mirror neutrinos, being Standard Model singlets, can play the role of
sterile neutrinos and one has a natural way to
understand why a singlet fermion,
$\nu_s$, may have such a tiny mass.

As mentioned before, the scale of electroweak
symmetry
breaking in the mirror sector $v'$ is assumed to be different from that in
the familiar sector ($v$)\cite{bere}. This is achieved by incorporating a
mirror odd field
$\eta$ which has a VEV and asymmetrizes the $SU(2)$ doublet Higgs mass terms
in both sectors. (Equivalently, this can be achieved by the soft breaking
of the mirror symmetry.) Due to different weak scales, the mirror neutrinos will
have different masses compared to the Standard Model ones.  If there is mixing
between the neutrinos of the two sectors, which is allowed by gauge symmetry,
then oscillations between the two types of neutrinos will occur.

Assume that our world respects $L_e+L_\mu-L_\tau$ and the mirror
world respects the analogous $L_e'+L_\mu'-L_\tau'$.  If these
symmetries were exact, a linear combination of familiar neutrinos, $\nu_e$
and $\nu_{\mu}$, as
well as another one of mirror neutrinos, i.e., $\nu_e'$ and
$\nu_\mu'$, will remain massless.  We shall identify the second linear
combination
as $\nu_s$.  Such global symmetries are perhaps good only up to quantum
gravitational corrections. We assume that they are indeed broken
by Planck scale effects. Therefore, once the Planck
scale breaking terms are included in the Lagrangian, both the massless
neutrino states would pick up small
masses. They give a mass to $\nu_s$ of order $\delta m_0 \sim v'^2/M_{\rm
Pl} $. If we choose the mirror weak scale, $v' \sim
(2-4)$ TeV, one finds $m_{\nu_s} \sim (2-8) \times 10^{-3}$ eV.
There will be Planck suppressed corrections to other entries in the
neutrino mass matrix as well, but they will not be of much consequence for
our purpose.  For example, the $\nu_e\nu_e$ entry will be corrected
by a negligible amount $\sim 10^{-5}$ eV.

Let us now discuss how the dominant entries of the neutrino mass matrices
arise. In the familiar sector,
the dominant entries of $M_\nu$, in units of $m_0$,  are the 1
and the $a$ entries.
The 1 arises via the seesaw mechanism that we describe now.

It proves convenient to use the type II seesaw
formula \cite{type2} for this purpose, where the tiny VEV of a $B-L=2$
triplet, rather than a heavy Majorana neutrino, gives small mass to the
neutrino. To implement this mechanism\cite{type2},
we introduce into the Standard Model in both sectors, $SU(2)$
triplets $\Delta_L$ and $\Delta'_L$ with hypercharges $+2$.
These fields will have the following Yukawa couplings to the lepton
doublets:
\begin{equation}
{\cal L}_{\rm YUK} = f_{ij}(L_i^T i\tau_2 \vec{\tau}.\vec{\Delta}_LL_j
+ L_i^{\prime T} i\tau_2 \vec{\tau}.\vec{\Delta}_L^\prime
L^\prime_j )+ h.c.
\end{equation}
By itself, Eq. (14) does not break lepton number, but the $\Delta_L$ and
$\Delta'_L$ fields will also have gauge invariant couplings to the Standard
Model Higgs doublet fields that violate lepton number.
The origin of the naturally small VEVs can be seen from the
following
effective potential, where mirror symmetry is assumed to be softly broken:
\begin{eqnarray}
V(\phi,\phi',\Delta_L,\Delta'_L)~=~+M^2(\Delta^{\dagger}_L\Delta_L
+ {\Delta'}^{\dagger}_L\Delta'_L) +
(\mu\Delta_L^\dagger\phi\phi+\mu'\Delta^{\prime \dagger}_L\phi'\phi') +.....
\end{eqnarray}
Here the ... denote higher order polynomial terms that are not relevant
for the present discussions.
The scales in Eq. (15) is chosen to be high seesaw scales in both the
sectors. The VEVs of the triplet fields are given by
\begin{equation}
v_T\equiv \left\langle \Delta^0_L\right\rangle \simeq
\frac{\mu v^2}{M^2};~~~
v'_T\equiv \left\langle \Delta^{0 \prime}_L\right\rangle \simeq
\frac{\mu'v^{\prime 2}}{M^2}~.
\end{equation}
The value of  $M^2/\mu$ is fixed so that the active species has
a mass term (connecting $\nu_\mu$ and $\nu_\tau$) of order 1 eV
(the 1 entry).  That sets $M^2/\mu \sim 3 \times 10^{13}$ GeV.
The mirror seesaw scale $M^2/\mu'$ is not identical,
we shall choose it to satisfy the cosmological requirement that
during big bang nucleosynthesis, no more than one sterile neutrino
species be in thermal equilibrium with the photons. This requires
$M^2/\mu'\simeq (10^6-10^{7})$ GeV.  The heavy mirror neutrinos
will then have a degenerate mass of order $(10-100)$ GeV.

The mixing term $am_0$
of Eq. (1) is also of order 1 eV.  This entry involves the mixing of $\nu_e$
with $\nu_s$ and thus will require the
breaking of the
two $L_e+L_\mu-L_\tau$ symmetries to a single $U(1)$.
This is achieved by
choosing a Higgs field $\chi$, which transforms like a bidoublet,
i.e., $(2,1; 2',1')$ under the two gauge groups and
 carries charge of $(-1,-1)$ under the two leptonic $U(1) \times U(1)'$
symmetries.
Its VEV will break the $U(1) \times U(1)'$ to the diagonal $U(1)$ subgroup
$\{(L_e+L_\mu-L_\tau)-
(L_e'+L_\mu'-L_\tau')\}$.  This is equivalent to $L_e+L_\mu-L_\tau-L_s$
since two of the mirror states are heavy.

The $\chi$ field has couplings to the lepton doublets of type
$L_e\chi
L'_e$. The field $\chi$ also acquires a VEV via seesaw type coupling like
the triplets, i.e., $V \supset \mu''\chi^\dagger \phi\phi'+
M^2_{\chi}\chi^{\dagger}\chi$.
Since $a m_0 \sim 1$ eV, we need
$\left\langle \chi \right\rangle \sim \frac{\mu''vv'}{M^2_{\chi}}\sim 1$ eV.
This can be obtained by choosing
$v\sim 200$ GeV, $v' \sim 2$ TeV, $\mu'' \sim M_{\chi} \sim
10^{14}$ GeV, so that $am_0\sim 0.4$ eV.
As for the entries $\epsilon_{1,2}$, they will arise in a manner similar
to the entries $1$ and $a$ except that their magnitudes
have to be explained by some effective Yukawa couplings being
order $0.02$.  As discussed before,
to get a nonzero value of $\delta \simeq (2 \times 10^{-3}- 3 \times
10^{-2})$, we invoke the Planck scale induced $U(1)'$ violating
couplings and $v' \simeq 4$ TeV.  This is about 25 times larger than $v$.
So cosmology of mirror sector will be identical to what has been
studied already\cite{teplitz}.

Since this model has two heavy sterile neutrinos (with masses in the multi
GeV range), one must address their cosmological implications. Their weak
interaction has a strength of $G_F\epsilon \simeq 10^{-3}G_F$ due to the higher
mirror symmetry breaking scale.
From this, one can calculate their decoupling temperature of the heavy
mirror neutrinos by
using the out of equilibrium condition as follows. The mirror neutrino
interaction rate is given by $G^2_F \epsilon^2
M^2(T'M)^{3/2}e^{-\frac{M}{T'}}$ where $T'$ is the temperature of the
mirror sector when the temperature of the visible sector is $T$, and
$M$ is the mass of the heavy neutrino. Let us
define the ratio of these temperatures to be $\beta = T'/T$. The out of
equilibrium condition then gives the following equation:
\begin{eqnarray}
\frac{n_{\nu'}(T_D)}{n_{\gamma}(T_D)} \simeq
\frac{\sqrt{g^*}x_D}{M_{P}G^2_F\epsilon^2M^3}~.
\end{eqnarray}
Here $x_D$ stands for the ratio $M/T_D$ where $T_D$ is
the decoupling temperature. $x_D$ is determined by the following equation:
\begin{eqnarray}
G^2_F\epsilon^2\beta^3M^3 x^{1/2}_De^{-\frac{x_D}{\beta}} \simeq
\frac{\sqrt{g^*}}{M_P}~.
\end{eqnarray}
This leads to $x_D \sim 3.6$. Using this, we see that for $M=100$ GeV,
the contribution of mirror heavy neutrinos to the energy density at the
BBN epoch is
$\frac{\rho_{\nu'}}{\rho_{\nu}}=\frac{n_{\nu'}(T_D)M}{n_{\gamma}(T_D)T_{BBN}}
\simeq 0.036$. Thus, even with both the heavy neutrinos, the total
contribution to the energy density during nucleosynthesis is negligible.
The lighter mirror neutrino will be in equilibrium during nucleosynthesis,
so the model predicts $N_\nu=4$ for nucleosynthesis.

The heavy mirror neutrinos are not stable, they
decay dominantly into the light mirror neutrino and
a mirror photon.  The decay lifetime can be estimated to be
$\tau^{-1}(\nu' \rightarrow \nu'_l+\gamma') \simeq
(\alpha/2)m_{\nu }^{\prime 3}[3 \epsilon G_F m_{\nu}^{\prime}m_\tau^2/(32
\pi^2 m_W^2)]^2 \simeq 10^{-4}$ sec. for $m_{\nu }' \sim 100$ GeV.
This will help eliminate most of the heavy neutrinos from the universe.

The mirror universe hypothesis is one possible framework that explains
naturally the existence of a light sterile neutrino.  Other approaches
based on radiative neutrino mass generation mechanisms \cite{zee} may
also provide realizations of our mass matrix, Eq. (1).

\section{Conclusions}

We have presented a simple ansatz for the four neutrino mass matrix
which leads to a mixed 2+2 scenario that fits
all neutrino oscillation data (solar, atmospheric and LSND).
The mass matrix follows from 
$L_e+L_\mu-L_\tau-L_s$ symmetry ($L_i$ is the $i$th lepton number).
When this symmetry is broken by the small mass of $\nu_s$, maximal
mixings required for explaining the solar and atmospheric neutrino
data result.  We have derived certain sum rules for neutrino
oscillation parameters within this scheme.  These sum rules are
consistent with present data.
The model predicts a small but nonzero value for the effective
$U_{e3}$ entry of the leptonic mixing matrix, which is approximately
equal to the
LSND oscillation angle, $\sim (0.02-0.03)$.  This prediction
can be used to test the model. We have also presented a gauge theory
realization of the neutrino mass matrix based on the mirror sector
hypothesis.  This specific realization leads to interesting dark matter
cosmology that has been studied previously.

\section*{Acknowledgements}

One of us (RNM) would like to thank D. Caldwell for discussions.
The work of KSB has been  supported in part by DOE Grant
\# DE-FG03-98ER-41076, a grant from the Research Corporation,
DOE Grant \# DE-FG02-01ER45684 and by the OSU Environmental
Institute. RNM is supported by the National Science Foundation
Grant No. PHY-0099544.

\end{document}